\documentclass[10pt,twocolumn,letterpaper]{article}
\usepackage{indentfirst}
\usepackage{colortbl} 
\usepackage{amssymb}
\usepackage{pifont}

\newcommand{\cmark}{\ding{51}} 
\newcommand{\xmark}{\ding{55}} 
\usepackage{cvpr}              

\usepackage[dvipsnames]{xcolor}

\definecolor{cvprblue}{rgb}{0.21,0.49,0.74}
\usepackage[pagebackref,breaklinks,colorlinks,citecolor=cvprblue]{hyperref}

\def\paperID{*****} 
\def\confName{CVPR}
\def\confYear{2024}

\title{HGNET: A Hierarchical Feature Guided Network \\for Occupancy Flow Field Prediction}

\author{Zhan Chen\textsuperscript{1}\quad Chen Tang\textsuperscript{1}\quad Lu Xiong\textsuperscript{1}\\
\textsuperscript{1}Tongji University\\
{\tt\small \{zhan\_chen, chen\_tang, xiong\_lu\}@tongji.edu.cn}
}

\begin{document}
\maketitle
\begin{abstract}
Predicting the motion of multiple traffic participants has always been one of the most challenging tasks in autonomous driving. The recently proposed occupancy flow field prediction method has shown to be a more effective and scalable representation compared to general trajectory prediction methods. However, in complex multi-agent traffic scenarios, it remains difficult to model the interactions among various factors and the dependencies among prediction outputs at different time steps. In view of this, we propose a transformer-based hierarchical feature guided network (HGNET), which can efficiently extract features of agents and map information from visual and vectorized inputs, modeling multimodal interaction relationships. Second, we design the Feature-Guided Attention (FGAT) module to leverage the potential guiding effects between different prediction targets, thereby improving prediction accuracy. Additionally, to enhance the temporal consistency and causal relationships of the predictions, we propose a Time Series Memory framework to learn the conditional distribution models of the prediction outputs at future time steps from multivariate time series. The results demonstrate that our model exhibits competitive performance, which ranks 3rd in the 2024 Waymo Occupancy and Flow Prediction Challenge.
\end{abstract}   

\section{Introduction}
\label{sec:intro}

Predicting the motion of multiple traffic participants has consistently been a significant challenge in autonomous driving technology. An accurate and robust prediction module must effectively handle a wide range of traffic scenarios and participant behaviors. Additionally, it is crucial to account for the potential interactions among different traffic participants, as simplistic predictions can result in unrealistic and contradictory outputs. Leveraging the robust capabilities of deep learning, recently proposed occupancy flow field prediction method offers an enhanced and more efficient representation for multimodal predictions in multi-agent scenarios~\citep{mahjourian2022occupancy}.

\begin{figure}[t]
  \centering
    \includegraphics[width=0.5\textwidth]{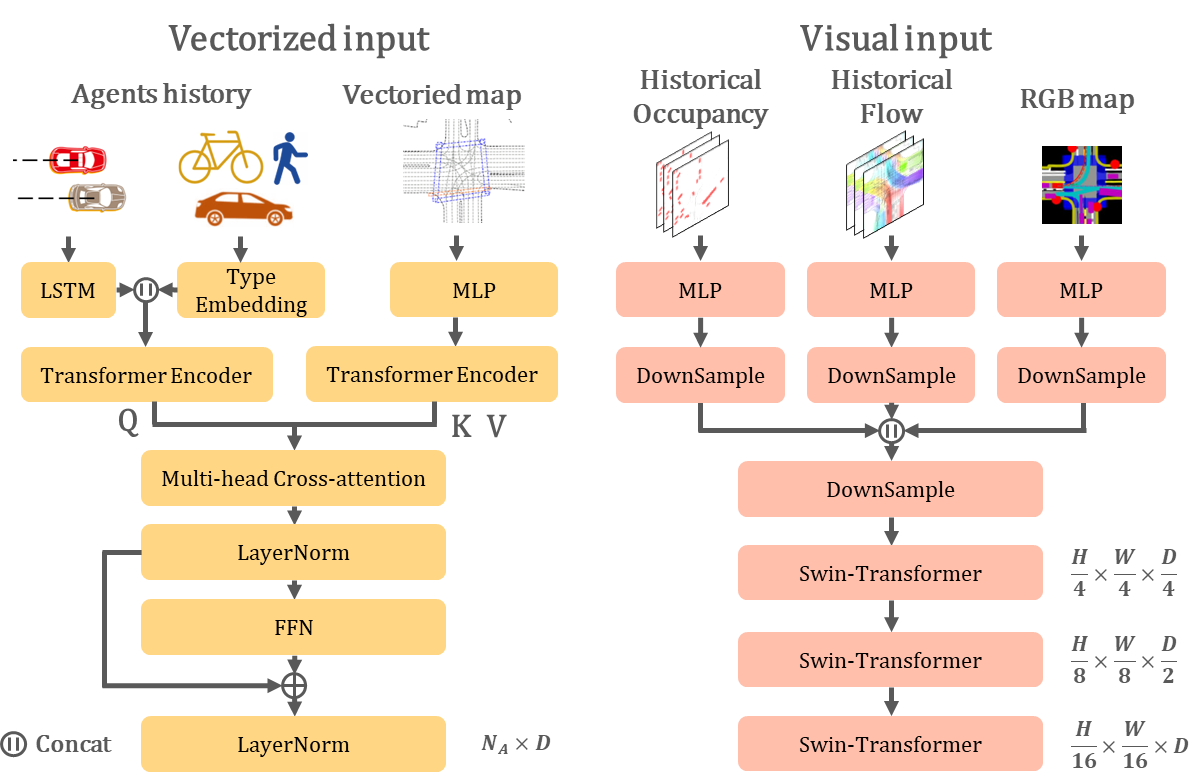}
   \caption{Transformer-based encoder for multimodal inputs.}
   \label{fig:encoder}
\end{figure}

However, current prediction techniques for occupancy flow field face several significant challenges. Firstly, while predictions for visible agent motion typically perform well within conventional tasks, forecasts for occluded obstacles often exhibit suboptimal performance. Additionally, there is a notable absence of suitable inference structures in network design, with many methods relying on high-dimensional abstract features at the network's front end to generate the final prediction targets~\citep{hu2022hope, liu2023multi}. Secondly, there is a lack of integration and correlation among the predictions for future flow and the future occupancy predictions for both visible and occluded obstacles. Effectively leveraging these correlations could substantially enhance the prediction performance of each component. Thirdly, for sequential prediction tasks, modeling the relationships between outputs at different time steps is essential for improving prediction accuracy.

In this technical report, we propose a hierarchical feature guided network for predicting occupancy flow field, along with several specialized structural designs to efficiently extract key features for forecasting the behaviors of multiple agents in complex traffic scenarios characterized by strong interaction relationships. Firstly, we employ a transformer-based encoder to extract visual and vectorized historical information, as well as map information, serving as input context tokens. Secondly, with the proposed Feature-Guided Attention (FGAT) module, we introduce a flexible and hierarchical framework that fully exploits the intrinsic relationships among flow, visible, and occluded agents' occupancy grid, thereby efficiently extracting correlated features. Thirdly, to extract the temporal relationships within prediction results over the forecast horizon and enhance the continuity and correlation among features, we have designed a Time Series Memory framework to capture and store temporal information. It should be noted that our proposed transformer-based prediction framework exhibits significant scalability and flexibility while ensuring superior predictive performance. Experiments on the Waymo Open Motion Dataset~\citep{ettinger2021large} demonstrate that HGNET can accurately forecast trajectories in the form of occupancy flow field at the scene level.


\section{Approach}

\begin{figure*}[htbp]
\centering
\includegraphics[width=\textwidth]{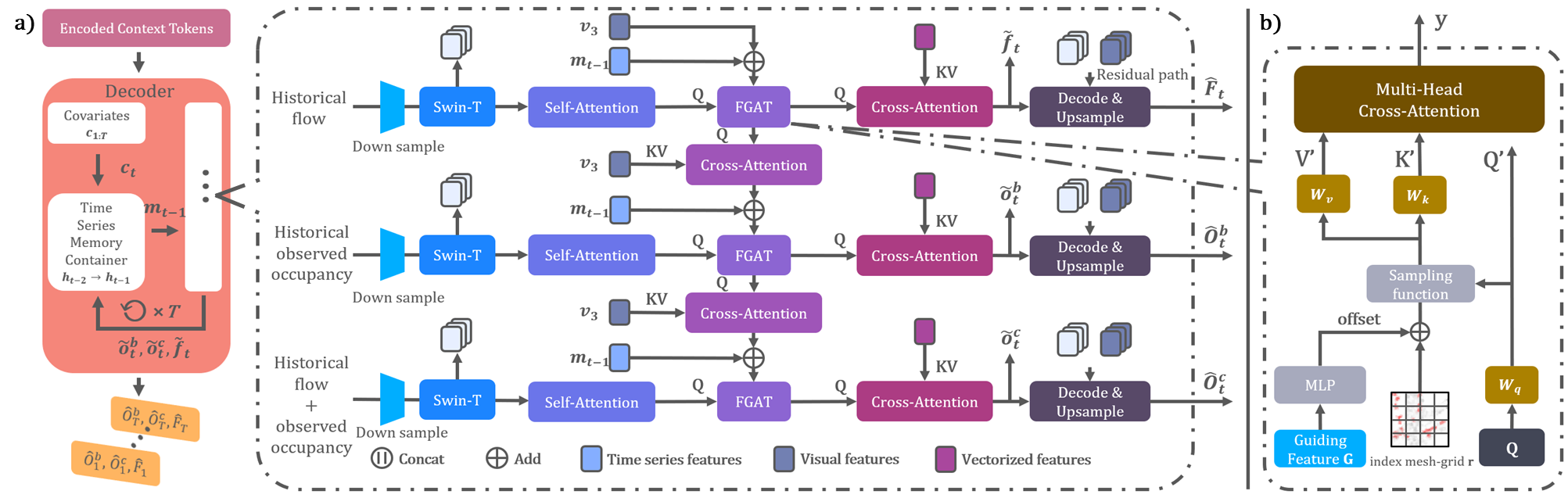}
\caption{\label{fig:decoder}a) Framework of the decoding pipeline. b) Structure of the Feature-Guided Attention module.
}
\end{figure*}

In this section, we provide a detailed introduction to the HGNET framework. First, we briefly introduce the network's input and encoder, with the overall architecture illustrated in Figure \ref{fig:encoder}. Next, we describe the proposed decoder, along with several specialized structures designed specifically for the prediction tasks. Finally, we explain the training objectives used to optimize the prediction model for joint occupancy flow field (OFF).

\subsection{Multi-modal Context Tokens Encoding}

To maximize the utilization of available multimodal input information, we employ two types of input information, including vectorized input and visual input. Vectorized input consists of the historical trajectory state sequences of $N_A$ agents within the scene over the past $T_h$ time steps, containing information including position, velocity, heading angle, and agent type. We also introduce vectorized map information $\mathcal{M}_{vec}$ to the system. Finally, the positional attributes of all agents and map elements are transformed into the local coordinate system of the ego vehicle. For visual input related to the prediction task of occupancy flow, we establish a historical occupancy grid along with the backward flow field between time steps $t=-T_h$ and $t=0$. Additionally, following the approach in~\citep{liu2023multi}, we introduce an RGB visualization representation $\mathcal{M}_{vis}$ of the map network to thoroughly incorporate essential map information like traffic light signals.

As shown in Fig. \ref{fig:encoder}, to consider the interaction among all elements within the traffic scenario, the historical states of all agents are firstly encoded by LSTM networks for all traffic agents, and concatenate it to agent's type embedding output. Then a two-layer self-attention Transformer encoder is applied to model the agent-agent interaction. The vectorized map waypoints are effectively encoded by a MLP layer as the latent feature, followed by a self-attention Transformer encoder. To capture relationships and dependencies between agents and map, we employ a cross-attention Transformer as agent-map interaction encoder, utilizing agent's interaction feature as query ($\mathbf{Q}$) and map feature encoded from vectorized map as key and value ($\mathbf{K, V}$). Without loss of generality, we let all latent features have $D$ hidden dimensions. Therefore, the vectorized tokens have the shapes of $[N_A, D]$.
For visual features, the original inputs are initially encoded by three MLP layers, then down sampled separately. We concatenate them with each other as a whole visual feature and feed it into the Swin-Transformer-based encoder. Each Swin-Transformer module comprises a two-layer Transformer equipped with both window self-attention and shifted window self-attention. This configuration facilitates comprehensive interaction modeling for visual features through global and intersected attention mechanisms. Additionally, each attention module incorporates multi-head attention with relative positional bias. The outputs from three stages of Swin-Transformer blocks are aggregated into a list $\mathbf{v}_1, \mathbf{v}_2, \mathbf{v}_3$ with shapes of $[\frac{H}{4},\frac{W}{4},\frac{D}{4}], [\frac{H}{8},\frac{W}{8},\frac{D}{2}], [\frac{H}{16},\frac{W}{16}, D]$ respectively, and serve as the final output of visual features.

\subsection{Hierarchical Feature Guided Decoder}
To organize the prediction inference sequence of various prediction targets more systematically and fully leverage the guiding role of different features, we design the structure of the hierarchical feature guided decoder as shown in Fig. \ref{fig:decoder}. We choose flow as the first prediction and utilize its high-dimensional features to inform subsequent prediction tasks for it represents the changes in occupancy grids between adjacent timesteps. Though occluded occupancy cannot be directly inferred, the relevant features can be effectively extracted using visible information and historical data~\citep{shao2023reasonnet}. Thus, we predict occluded occupancy as the last prediction target, merging the features of both flow and observed occupancy as guiding features. For each prediction pathway, we first encode the corresponding inputs using a similar method as described before, obtaining a feature list with the same shape as the visual features (where the original features of occluded occupancy are derived from visible occupancy and flow). Subsequently, the encoded features pass through a self-attention layer and are fed into our proposed FGAT module as the query.

\noindent \textbf{Feature-Guided Attention module.}
We designed the FGAT module to amplify the query with corresponding features guided by learnable offsets generated from the guiding feature. Within the hierarchical network architecture, the FGAT module aggregates various features from future timesteps. Particularly, except for the top-level FGAT module, all guiding features are first input into a cross-attention module as queries (with visual feature $\mathbf{v}_3$ as keys and values) then added with time series feature $\mathbf{m}_{t-1}$, before entering the FGAT module. Given the encoded historical feature as query, guiding feature $\mathbf{Q}, \mathbf{G}\in \mathbb{R}^{H/16\times W/16\times C}$, and a uniform index mesh-grid of points $\mathbf{r}\in \mathbb{R}^{H/16\times W/16\times 2}$ as the references, the offsets $\Delta \mathbf{r}$ for reference points are generated from the guiding feature by a MLP layer along with a \textit{tanh} layer:

\begin{equation}
    \begin{aligned}\label{4}
    &\mathbf{Q'}=\mathbf{Q}W_q, \mathbf{K'}=\mathbf{x}W_k, \mathbf{V'}=\mathbf{x}W_v,\\
    &\mathbf{x}=f_\phi(\mathbf{Q'};\mathbf{r}+\Delta \mathbf{r}), \Delta \mathbf{r}=\texttt{tanh}(\texttt{MLP}(\mathbf{G})),
    \end{aligned} 
\end{equation}

\noindent where $\mathbf{K'}$ and $\mathbf{ V'}$ represent the feature-guided key and value embeddings, and we use a bilinear interpolation as $f_\phi(\cdot;\cdot)$:

\begin{equation}\label{5}
    f_\phi(\mathbf{G;R})=\sum_{(x,y)} g(\mathbf{R}_x,x)g(\mathbf{R}_y,y)\mathbf{G}[y,x,:],
\end{equation}

\noindent where $g(i,j)=\texttt{max}(0,1-|i-j|)$ and $(x,y)$ represents every point location of $G\in \mathbb{R}^{H/16\times W/16\times D}$. Finally we perform a multi-head cross-attention on $\mathbf{Q'}, \mathbf{K'}, \mathbf{V'}$ with relative positional bias $B$, the projection matrice $W_o$ and the dimension of the key token $d$,

\begin{equation}
    \begin{aligned}\label{6}
    &\texttt{MHCA}(\mathbf{Q',K',V'}) = (h_i||...||h_{\mathbf{M}})W_o,\\
    &h_i = \texttt{softmax}(\mathbf{Q'K'}^\top/\sqrt{d}+B)\mathbf{V'},
    \end{aligned} 
\end{equation}

\noindent \textbf{Time Series Memory Framework.} To improve the accuracy of temporal feature prediction results, we adopt this framework to learn a model of the conditional distribution of future time steps of a multivariate time series  given the historical features and covariates as:

\begin{equation}\label{7}
    q(\mathbf{y}_{t_0:T}|\mathbf{y}_{1:t_0-1},\mathbf{c}_{1:T})=\prod_{t=t_0}^Tq(\mathbf{y}_t|\mathbf{y}_{1:t-1},\mathbf{c}_t),
\end{equation}

\noindent where $t_0$ denotes the current prediction time step. We use the embeddings of future time steps as covariates $\mathbf{c}_{1:T}$. To model the temporal dynamics via the updated hidden state $h_{t-1}$, we employ three multi-layer GRU networks to encode the time series sequence up to time step $t-1$, given the covariates of the next time step $\mathbf{c}_t$:

\begin{equation}\label{8}
    \mathbf{m}_{t-1},\mathbf{h}_{t-1}=\mathbf{GRU}(\texttt{concat}(\mathbf{y}_{t-1},\mathbf{c}_t),\mathbf{h}_{t-2}),
\end{equation}

\noindent where $\mathbf{h}_0=\mathbf{0}$ and $\mathbf{m}_{t-1}$ is the output of GRU network. In three prediction heads, $\mathbf{y}_{t-1}$ represents $\Tilde{\mathbf{f}}_{t-1}, \Tilde{\mathbf{o}}_{t-1}^b, \Tilde{\mathbf{o}}_{t-1}^c$ respectively. They are the outputs of the cross-attention module, where the output of the FGAT module serves as the query, the encoded vector features are used as the key and value. By dynamically updating the hidden states, the information from previous time steps is preserved and fused for predicting features at the next time step.

Finally, we decode the flow and occupancy from the feature tensors using feature pyramid network (FPN), which consists of multi-layer 2D-CNNs and upsampling layers, along with additional 2D-CNNs employed to process the features in the residual paths.

\subsection{Training Objectives}
For the occupancy loss $\mathcal{L}_{occ}$, we utilize the focal loss and the cross-entrophy loss for the observed and occluded occupancy regression. Similar to~\cite{mahjourian2022occupancy}, smooth L1 loss is applied as flow loss $\mathcal{L}_{f}$ to supervise the flow prediction. The final multi-task training objective sum up the loss terms scaled by the size of the grid map (with height $h$ and width $w$) and length of timesteps of the output:
\begin{equation}\label{9}
    \mathcal{L}=\frac{1}{hwT}(100\mathcal{L}_{occ}+\mathcal{L}_f)
\end{equation}

\begin{table*}[]
\centering
\begin{tabular}{l|cc|cc|c|cc}
\specialrule{1pt}{0pt}{0pt}
\multicolumn{1}{c|}{\rule{0pt}{2.5ex} Evaluation Metrics \rule[-1.5ex]{0pt}{0pt}} & 
\multicolumn{2}{c|}{\rule{0pt}{2.5ex}Observed Occupancy\rule[-1.5ex]{0pt}{0pt}} & 
\multicolumn{2}{c|}{\rule{0pt}{2.5ex}Observed Occupancy\rule[-1.5ex]{0pt}{0pt}} & 
\multicolumn{1}{c|}{\rule{0pt}{2.5ex}Flow\rule[-1.5ex]{0pt}{0pt}} & 
\multicolumn{2}{c}{\rule{0pt}{2.5ex}Flow-grounded Occupancy\rule[-1.5ex]{0pt}{0pt}} \\ \hline
\textbf{Method}    & \textbf{AUC} $\uparrow$      & \textbf{Soft-IoU} $\uparrow$   & \textbf{AUC} $\uparrow$      & \textbf{Soft-IoU} $\uparrow$   & \textbf{EPE} $\downarrow$              & \textbf{AUC} $\uparrow$       & \textbf{Soft-IoU} $\uparrow$    \\ \hline
DOPP               & \textbf{0.797}    & 0.343               & \textbf{0.194}    & 0.024               & \textbf{2.957}            & \textbf{0.803}      & \textbf{0.516}        \\
STNet              & 0.755             & 0.230               & 0.166             & 0.018               & 3.378                     & 0.756               & 0.443                 \\ \hline
\rowcolor[HTML]{D5D5D5} 
\textbf{Ours}      & 0.733             & \textbf{0.421}      & 0.166             & \textbf{0.039}      & 3.670                     & 0.740               & 0.450                 \\ \specialrule{1pt}{0pt}{0pt}

\end{tabular}
\caption{Summary of the testing performance on the Waymo occupancy and flow prediction benchmark.}
\label{tab:1}
\end{table*}

\begin{table}[]
\begin{tabular}{lc|cccc}
\specialrule{1pt}{0pt}{0pt}
\multicolumn{1}{c}{\begin{tabular}[c]{@{}c@{}}FG-\\ AT\end{tabular}}                        & \begin{tabular}[c]{@{}c@{}}time  \\ series\end{tabular} & \begin{tabular}[c]{@{}c@{}}Observed\\ AUC$\uparrow$\end{tabular} & \begin{tabular}[c]{@{}c@{}}Occluded\\ AUC$\uparrow$\end{tabular} & \begin{tabular}[c]{@{}c@{}}Flow\\ EPE$\downarrow$\end{tabular} & \begin{tabular}[c]{@{}c@{}}FG\\ AUC$\uparrow$\end{tabular} \\ \hline
\xmark                        & \xmark                                                      & 0.713                                                  & 0.131                                                  & 3.905                                              & 0.719                                            \\
\cmark                         & \xmark                                                         & 0.721                                                  & 0.154                                                  & 3.724                                              & 0.733                                            \\ \hline
\cmark & \cmark                                 & \textbf{0.742}                                         & \textbf{0.158}                                         & \textbf{3.561}                                     & \textbf{0.743}                                   \\ \specialrule{1pt}{0pt}{0pt}
\end{tabular}
\caption{Ablation study on FGAT module and time series memory framework.}
\label{tab:2}
\end{table}

\section{Experiments}
\subsection{Implementation Details}
The hidden feature dimension is 256, We choose GELU as the activation function in all encoders and RELU in the decoder. Dropout is followed after every MLP layer, all with a dropout rate of 0.1. we use a distributed training strategy on 2 Nvidia RTX 6000 Ada GPUs with a total batch size of 16. The training process lasts 16 epochs. We use the Adam optimizer  for training with the initial learing rate of 1e-4, and the learning rate decayes by a factor of 50\% every 2 epochs.

\subsection{Quantitative Results}
The performance of HGNET on the the Waymo occupancy and flow prediction benchmark is shown in Tab.~\ref{tab:1}, where we can see that our proposed approach outperforms other method on some metrics, and overall, it demonstrates good performance across other metrics and exhibits a certain level of competitiveness. 

\subsection{Ablation Study}
We conduct an ablation study to investigate the infuences of key modules in our proposed method, i.e., FGAT module and time series memory framework. We conducted experiments with two ablation variants of our model: one excluding the time series memory framework (i.e., without updates to the hidden states of the time series and the fusion of time series features) and another excluding the FGAT module (substituting it with a standard cross-attention module). As shown in Tab.~\ref{tab:2}, the performance metrics on the Waymo occupancy flow validation set exhibit a decline across all metrics for the ablated models. These results substantiate the efficacy of our proposed framework in enhancing prediction accuracy.


\section{Conclusion}
We propose HGNET, a hierarchical multi-modal feature guided framework for joint multi-agent occupancy flow field prediction. Leveraging the proposed Feature-Guided Attention module for feature guidance and an effective Time Series Memory framework for temporal feature extraction, our model achieves accurate multi-agent motion prediction in the form of occupancy flow fields. Experimental results demonstrate that our method achieves competitive performance on the Waymo occupancy and flow prediction benchmark.

{
    \small
    \bibliographystyle{ieeenat_fullname}
    \bibliography{main}
}


\end{document}